\documentclass{aa}  

\usepackage{graphicx}
\usepackage{txfonts}
\usepackage{natbib}
\usepackage[flushleft]{threeparttable}
\newcommand{\PSO}{PSO~J0309+27 }

\begin{document}

   \title{Direct observation of an extended X-ray jet at $z$=6.1}

   \author{L.~Ighina
          \inst{1,2}
          \and
          A.~Moretti\inst{1}
          \and 
          F.~Tavecchio\inst{3}
          \and
          A.~Caccianiga\inst{1}
          \and
          S.~Belladitta\inst{1,2}
          \and
          D.~Dallacasa\inst{4,5}
          \and
          R.~Della Ceca\inst{1}
          \and
          T.~Sbarrato\inst{3}
          \and
          C.~Spingola\inst{5}
          }

   \institute{
    INAF -- Osservatorio Astronomico di Brera, via Brera 28, 20121, Milano, Italy \\
              \email{lighina@uninsubria.it}
    \and
    DiSAT -- Universit\`a degli Studi dell'Insubria, via Valleggio 11, 22100 Como, Italy
    \and
    INAF -- Osservatorio Astronomico di Brera, via E. Bianchi 46, 23807 Merate, Italy
    \and
    Department of Physics and Astronomy, Università degli Studi di Bologna, Via Gobetti 93/2, 40129 Bologna, Italy
    \and
    INAF -- Institute for Radioastronomy, via Gobetti 101, 40129, Bologna, Italy
    \\
             }

   \date{Received November 16, 2021; accepted January 12, 2022}

 
  \abstract
   {
   
   We report on the direct observation of an extended X-ray jet in the $z$=6.1 radio-loud active galactic nucleus PSO~J030947.49+271757.31 from a deep Chandra X-ray observation (128 ksec). This detection represents the most distant kiloparsec-scale off-nuclear emission resolved in X-rays to date. The angular distance of the emission is $\sim$4\arcsec (corresponding to $\sim$20~kpc at $z$=6.1), along the same direction of the jet observed at parsec scales in previous VLBA high-resolution radio observations. Moreover, the 0.5-7.0~keV isophotes coincide with the extended radio emission as imaged by the VLA Sky Survey at 3 GHz. The rest-frame 2–10~keV luminosity of the extended component is  L$_{2-10keV}$=5.9$\times$10$^{44}$~erg~s$^{-1}$, about 8\% of the core: this makes it one of the most luminous jets resolved in X-rays so far.
   Through spectral energy distribution modelling we find that this emission can be explained by the inverse Compton interaction with the photons of the cosmic microwave background, assuming that the jet's physical parameters are similar to those in the local Universe. At the same time, we find that the radiation produced by a putative population of high-energy electrons through the synchrotron process observed at low redshift is quenched at high redshift, hence becoming negligible.

}
    
   \keywords{ Galaxies: active -- Galaxies: nuclei -- Galaxies: high-redshift -- Galaxies: jets -- quasars: general  -- X-rays: general -- individual: PSO~J030947.49+271757.31
               }

   \maketitle
%

\section{Introduction}
Active galactic nuclei (AGNs) are the brightest permanent astronomical objects and one of the most valuable sources of information from the early Universe. A fraction of them ($\sim$10-15\%; e.g. \citealt{Liu2021,Diana2021}) are able to expel part of the accreting matter in the form of two collimated relativistic jets originating very close to the supermassive black hole (SMBH) and extending even up to a few megaparsecs \citep[e.g.][]{Blandford2019}.

\
Understanding the mechanisms responsible for the launch and emission of these jets is of crucial importance for constraining the kinetic power they carry and therefore their feedback on the intergalactic medium as a function of redshift \citep[e.g.][]{Nesvedba2007, Fabian2012}. It is also crucial for constraining the cosmological evolution of the SMBHs hosted in jetted AGNs \citep[e.g.][]{Fabian2014} and the relative contribution of jets to the re-ionisation of the Universe at $z$\textgreater6 \citep[e.g.][]{Torres2020}.

Even though we are able to resolve the emission and the structure of relativistic jets up to milliarcsecond scales with radio observations (e.g. \citealt{Boccardi2017}), in X-rays we usually observe only the unresolved emission produced in the innermost compact region of the jet, in particular when it is closely aligned to our line of sight \citep[i.e. blazars; e.g.][]{Bhatta2018}. However, there are a few exceptions where, thanks to the angular resolution of the \textit{Chandra} X-ray telescope ($\sim$0.5\arcsec; \citealt{Weisskopf2000}), we are able to resolve the most extended regions of relativistic jets in the X-ray band and therefore study their properties even at high energies.

Nevertheless, more than 20~years after the direct detection and study of the first extended jet observed with the \textit{Chandra} telescope in X-rays \citep{Schwartz2000}, there is no homogeneous consensus on the mechanism responsible for the radiation observed several kiloparsecs away from the SMBH. One of the first and most popular interpretations proposed for the extended high-energy emission is the interaction of electrons with the photons of the cosmic microwave background (CMB) through the inverse Compton (IC) process \citep[IC/CMB; e.g.][]{Tavecchio2000, Celotti2001}. Despite this model's initial success, the non-detections of a strong and permanent $\gamma$-ray emission from sources in the local Universe ($z$\textless1) ruled out the possibility that this process is the main contributor to the observed X-ray emission in most objects \citep[e.g.][]{Meyer2015,Breiding2017}. At the same time, synchrotron emission from a second highly energetic population of electrons, different from the one responsible for the radio emission, seems to be favoured (\citealt{Harris2002,Jester2002,Georganopoulos2006}). 
In any case, we still expect the IC/CMB interaction to take place to a certain degree, especially at high redshift, thanks to the strong evolution of the CMB energy density, $\propto$(1+$z$)$^4$ \citep[e.g.][]{Ighina2021}. For this reason, the best way to characterise this interaction and its contribution to the overall X-ray emission is to focus on the high-$z$ jetted AGN population and then extrapolate the expected amount to lower redshifts \citep{Worrall2020}, where many extended jets have already been analysed (e.g. \citealt{Harris2006,Marshall2011}). At the same time, the IC/CMB radiation is also strongly dependent on the viewing angle (more so than the synchrotron radiation; e.g. \citealt{Worrall2009}) and can therefore be more easily observed in blazars, where the relativistic jet is oriented close to the line of sight, even if they are in the local Universe (e.g. \citealt{Meyer2019}).

In this context, the highest redshift flat-spectrum radio quasar (hereafter simply blazar) known to date, PSO~J030947.49+271757.31 (hereafter PSO~J0309+27; \citealt{Belladitta2020}), is the ideal candidate for searching for and studying the emission of jets on the kiloparsec scale in the primordial Universe. To this end, we performed relatively deep \textit{Chandra} X-ray observations (128~ksec; P.I. A. Moretti, Archive Seq. Num. 704032, 704242).
\PSO was discovered by combining the NRAO VLA Sky Survey (NVSS; at 1.4~GHz; \citealt{Condon98}) with the Panoramic Survey Telescope and Rapid Response System (Pan-STARRS; \citealt{Chambers2016}) and then confirmed spectroscopically \citep{Belladitta2020}. Based on the flat radio spectrum in the 0.15--1.4~GHz observed band and the high X-ray luminosity derived from a \textit{Swift}-XRT observation, this source was classified as a blazar, the first observed at $z$\textgreater6. 
Thanks to its very bright nature compared to other $z$\textgreater6 AGNs, \PSO has been the target of several observational campaigns aimed at constraining the full extent of its spectral energy distribution (SED; see \citealt{Belladitta2020,Spingola2020,Moretti2021, Belladitta2021} and also \citealt{Mufakharov2021}).
In this work we focus on the multi-wavelength characterisation of the emission extending a few arcseconds from the AGN position, which was revealed from dedicated X-ray \textit{Chandra} observations and represents the highest redshift jet resolved in X-rays currently known.

In Sect. \ref{sec:multi_data} we present the available radio and optical data that can constrain the arcsecond emission of \PSO as well as the X-ray imaging and spectral analysis of the \textit{Chandra} observations. In Sect. \ref{sec:discussion} we model the X-ray and radio measurements considering both the IC/CMB interaction and the synchrotron emission of a highly energetic population of electrons. We then compare currently available data for the extended jet of \PSO to other resolved jets at lower redshift reported in the literature. Finally, in Sect. \ref{sec:conclusions} we summarise our results and conclusions.\\

Throughout the paper we assume a flat $\Lambda$ cold dark matter cosmology with $H_{0}$=70 km s$^{-1}$ Mpc$^{-1}$, $\Omega_m$=0.3, and $\Omega_{\Lambda}$=0.7, where 1\arcsec\, corresponds to a projected distance of 5.66~kpc at $z$=6.1. Spectral indices are given assuming $S_{\nu}\propto \nu^{-\alpha}$, and all uncertainties are reported at 90\% confidence unless otherwise specified.


   \begin{figure}
   \centering
   \includegraphics[width=\hsize]{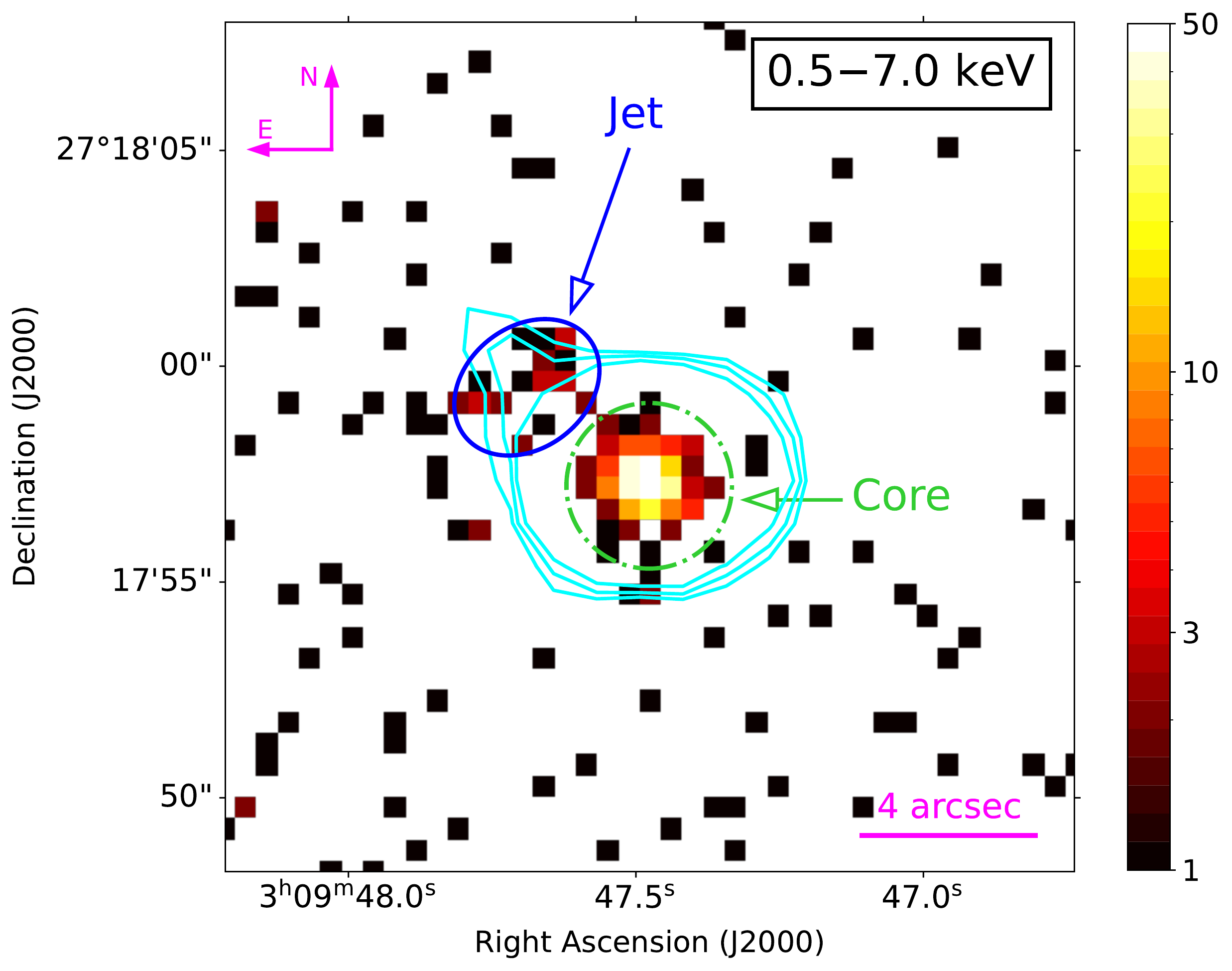}
   \caption{3~GHz radio contours from the VLASS survey (in cyan) overlaid on the 0.5--7~keV X-ray image of \PSO from the 128~ksec \textit{Chandra} observation. The count colour scale is displayed on the right. For display purposes, only the (3, 3$\sqrt{2}$, 6) $\times$ the off-source RMS (=0.12~mJy~beam$^{-1}$) radio contours are reported. The dashed green circle and the solid blue ellipse are the regions used for the extraction of the core and the jet X-ray spectra, respectively.}
   \label{fig:image_XR}%
    \end{figure}

\section{Multi-wavelength data}
\label{sec:multi_data}

In this section we describe the archival and recent proprietary data for \PSO relevant to the study of the source's jet at arcsecond resolution. A detailed discussion on the components at scales $\lesssim$1\arcsec \, is given by \cite{Spingola2020} in the radio, \cite{Belladitta2021} in the optical to near-infrared (NIR), and \cite{Moretti2021} in X-rays.

\subsection{Radio observations}
Besides the NVSS survey, \PSO was detected in the TIFR  Giant Metrewave Radio Telescope Sky Survey (TGSS; \citealt{Intema2017}) at 150~MHz  and in the Karl G. Jansky Very Large  Array Sky Survey (VLASS; \citealt{Lacy2020}) at 3~GHz. In both of these surveys an extended component is visible in the north-east (NE) direction. In the TGSS image, the extended structure is possibly observed up to $\sim$20\arcsec \, ($\sim$110~kpc), although the signal-to-noise ratio is low ($\sim$2; see Fig. 3 in \citealt{Belladitta2020}); in the VLASS image the extension is present $\sim$4\arcsec \, from the core with a significance of approximately five times the RMS: S$_\mathrm{peak}$=0.63~mJy~beam$^{-1}$ (RMS=0.12~mJy~beam$^{-1}$; see Fig. \ref{fig:image_XR}). In this work we considered this value for the radio emission of the kiloparsec-scale? jet by adding a further 10\% to the uncertainty since the VLASS quick-look image flux calibration may be less reliable for faint sources (see \citealt{Gordon2020} for more details).

Moreover, \PSO was the target of Very Long Baseline Array (VLBA) observations at 1.5, 5, and 8.4~GHz in April 2020. From these radio observations at milliarcsecond angular resolution, \cite{Spingola2020} discovered the presence of a jet that extends for about 500~pc in projection in the NE direction as well (magenta dashed line in Fig. \ref{fig:NIR_images}). At the same time, low-resolution observations (between 22\arcsec at 1.4~GHz and 1\arcsec\ at 40~GHz) of \PSO with the Jansky Very Large Array revealed that the overall radio spectrum of the source is relatively steep ($\alpha_{\rm r}\sim$1; \citealt{Spingola2020}), indicating that its total radio emission is not dominated by the innermost regions of the jet.

\subsection{Optical and near-infrared observations}
\begin{figure}
   \centering
   \includegraphics[width=\hsize]{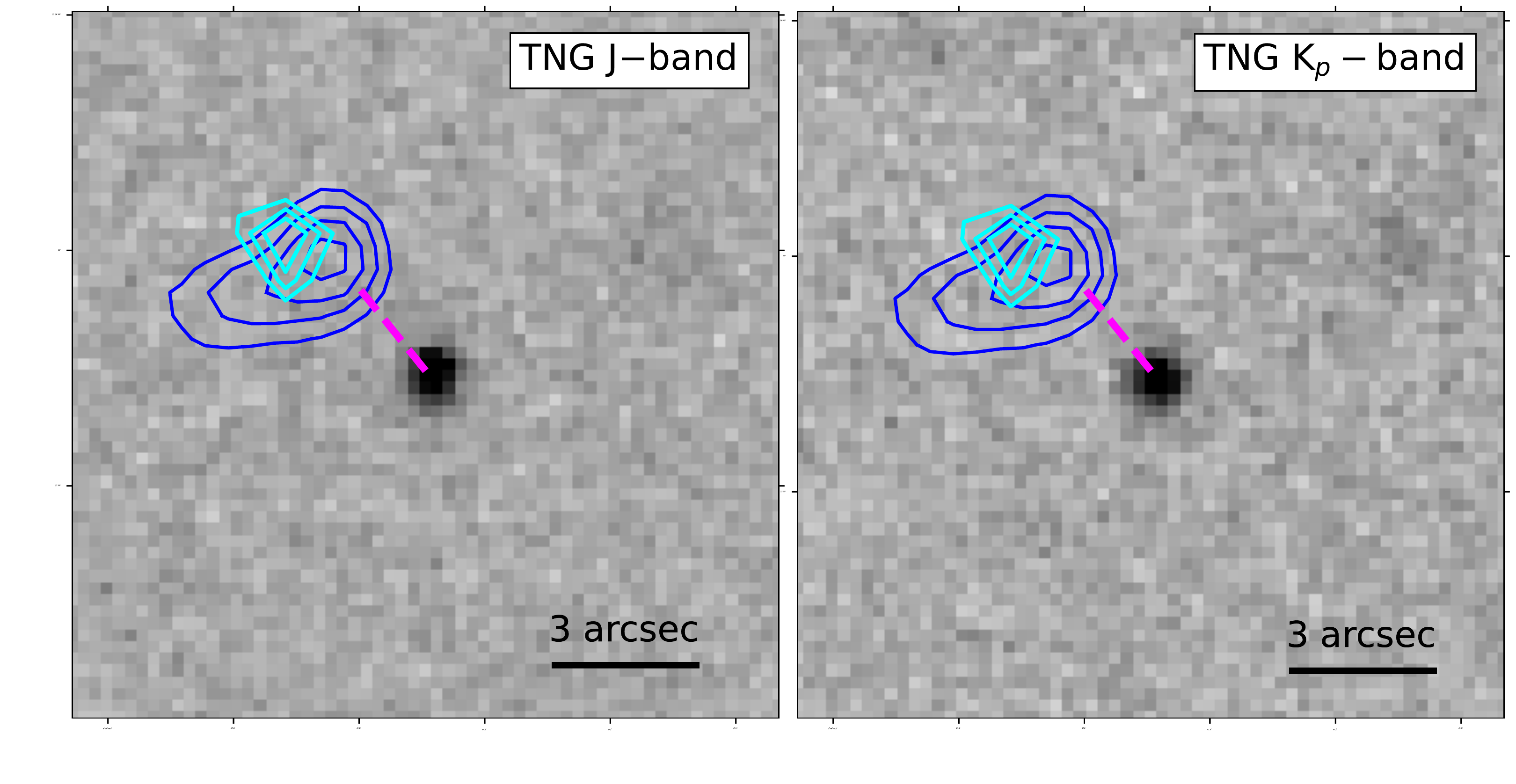}
      \caption{ $J$- and $K_{\rm p}$-band images of \PSO obtained with the TNG. The contours of the smoothed Chandra X-ray image, with the core's region subtracted, are shown in blue; the contours from the residual VLASS image  (i.e. with the core's emission subtracted) are displayed in cyan. The contour levels have been arbitrarily chosen to best show the jet and are always greater than three times the RMS of the given image. The dashed magenta line indicates the direction of the $\sim$500~pc radio jet described in \cite{Spingola2020}.}
         \label{fig:NIR_images}
   \end{figure}

In the optical and NIR bands, the field around \PSO was observed in the Pan-STARRS survey (in the $g$, $r$, $i$, $z$, and $Y$ filters) as well as through dedicated observations with the Telescopio Nazionale Galileo (TNG) in the $J$ and $K_\mathrm{p}$ filters (P.I. S. Belladitta; see Fig. \ref{fig:NIR_images} and \citealt{Belladitta2021} for further details). The corresponding 3$\sigma$ limiting magnitudes for an extended object of the size of the resolved X-ray photons (see Fig. \ref{fig:image_XR} and the next subsection) range from 23.2 to 21.2~mag. In all these observations, as well as in the Wide-field Infrared Survey Explorer Catalogue (CatWISE; \citealt{Eisenhardt2020}), no optical--IR counterpart associated with the extended jet is observed, suggesting that the observed X-ray emission is likely related to the jet of \PSO and not to a foreground or background source, such as an obscured active galaxy.

\subsection{Chandra X-ray imaging and spectral analysis}

\subsubsection{Astrometry}

\PSO has been observed for a total of 128~ksec with the \textit{Chandra} telescope (26.7~ksec in March 2020 and 101.7~ksec in November 2020). Data  reduction was performed using the \textit{Chandra} Interactive Analysis of Observations (\texttt{CIAO}) software  package  (v4.13; \citealt{Fruscinone2006}) with \texttt{CALDB} version  4.9.5.
The final image was reconstructed by correcting the astrometry of the single observation segments. To this end, we used the position of five objects detected both in Pan-STARRS and in the Chandra observation. The transformation matrix was produced through the \texttt{wcs\_match} task. We found typical shifts of +0.7 and +0.9~pixels (1 pixel=0.492\arcsec) in RA and Dec., respectively, with no significant rotation. This is not unexpected given that the overall 68\% uncertainty circle of the Chandra X-ray absolute position has a radius of 0.6\arcsec \footnote{\url{https://cxc.cfa.harvard.edu/cal/ASPECT/celmon/}}. 
Astrometry of the event file and images was modified accordingly by means of the  \texttt{wcs\_match} task. \PSO is detected with 320  photons in the 0.5-7.0~keV band within a 2\arcsec radius circle (98\% of the point spread function) with only $\sim$2.7 background photons expected. With the corrected astrometry, the position of the X-ray source is  RA=47.44786~deg and Dec.=+27.29922~deg with an error of 0.2\arcsec, which is mostly due to the residuals in the astrometric correction. This position is consistent with the Pan-STARRS optical and the VLASS/VLBA radio positions.

\subsubsection{X-ray imaging}

In Fig. \ref{fig:image_XR} we report the X-ray image obtained from the overall 128~ksec exposure in the 0.5--7~keV energy band with the radio contours at 3~GHz (from the VLASS survey) overlaid. 
In Fig. \ref{fig:NIR_images} we show the residual of the X-ray and radio images after the subtraction of the core emission (assumed to be point-like) overlaid on the NIR images. After the removal of the X-ray core, a significant excess of photons is present between 2\arcsec\ and 5\arcsec away from the position of PSO~J0309+27. Moreover, there is an overall positional agreement between the extended components in X-rays and in the radio, both directed towards the NE direction, similar to the position angle (P.A.) of the jet described in \cite{Spingola2020} at parsec scales.

We note that, even though radio data are too shallow for any detailed study, the jet seems to be resolved in X-rays with a lower surface brightness component towards the eastern direction. Such tangential extension could be related to a potential bend at kiloparsec scales. Indeed, if the jet is oriented close to our line of sight (as hinted by the multi-wavelength properties of PSO~J0309+27; e.g. \citealt{Spingola2020,Moretti2021}), even a small re-orientation of the jet would be amplified through projection and would result in a significant apparent change of direction. Similar bends have already been observed in the jets of several quasars, both at parsec (in the radio band; e.g. \citealt{Lister2021}) and kiloparsec scales (in the radio and X-ray bands; e.g. \citealt{Worrall2005,Marshall2011}). Therefore, it is likely that the overall extended X-ray emission of \PSO  visible in Fig. \ref{fig:NIR_images} is associated with different regions of the relativistic jet, before and after the bend. Nevertheless, deeper X-ray and radio observations at arcsecond resolution are needed to confirm this hypothesis and to fully characterise the morphology of this kiloparsec-scale jet. For simplicity, in the following we consider the entire extended emission to be produced in regions with similar physical properties.

\begin{figure}
   \centering
   \includegraphics[width=\hsize]{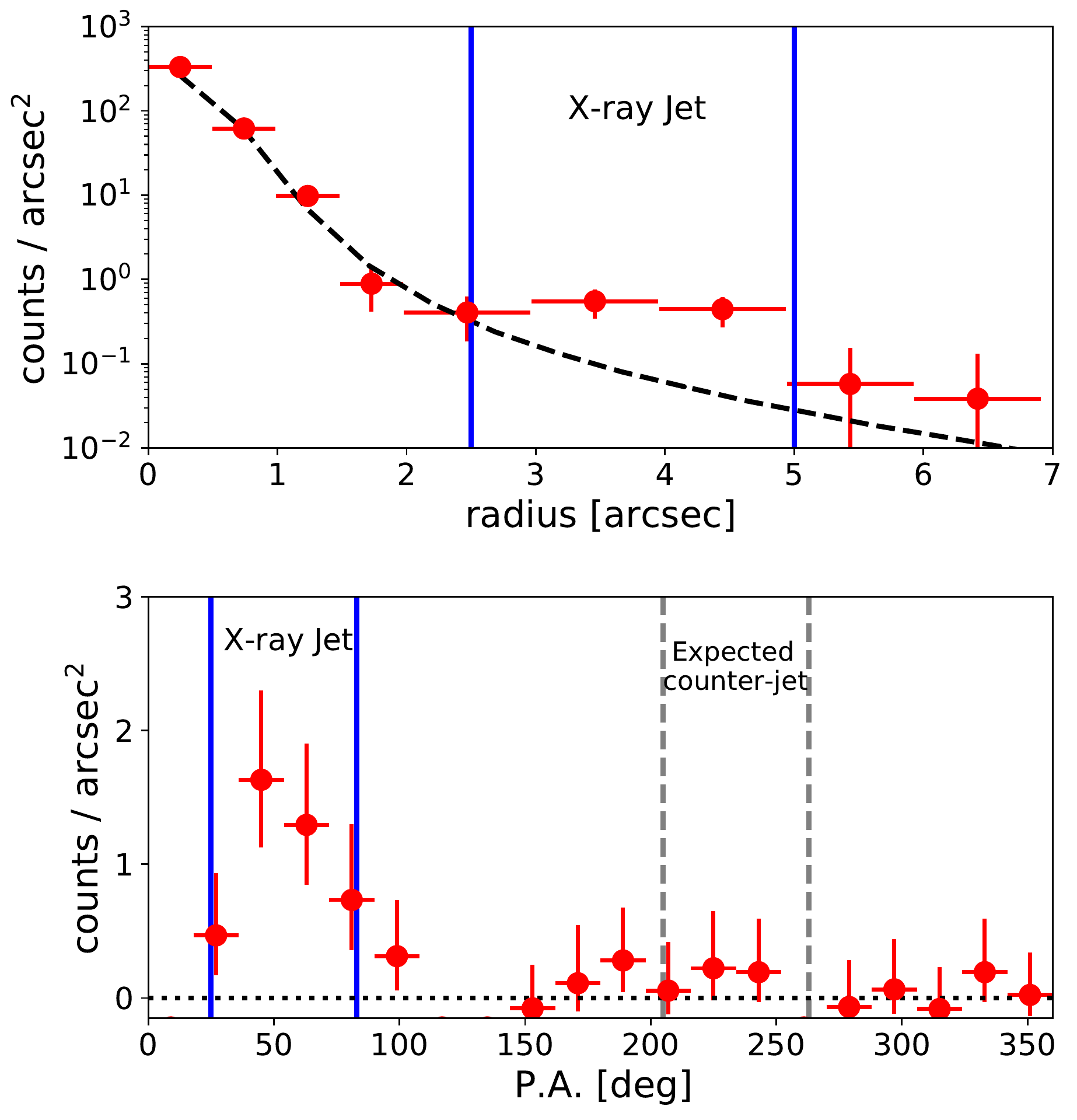}
      \caption{Observed radial (top) and angular (bottom) profiles of PSO~J0309+27 in the 0.5--7~keV energy band. In both cases the background has been subtracted and data points have Poissonian uncertainties. The dashed black line in the top panel is the expected PSF of the observation, computed with \texttt{MARX}. The vertical blue lines delimit the extraction region considered for the spectral analysis of the jet, and the area delimited by the grey lines is where the counter-jet is expected (see Sect. 2.3.2). The direction of the P.A. in the lower panel is east of north.}
         \label{fig:radial_prof}
   \end{figure}

In order to qualitatively constrain the  projected size of the emitting region, we started by computing the number of photons observed per unit of area around the core of \PSO as a function of distance. We show the obtained radial profile in Fig. \ref{fig:radial_prof}. We also report the expected distribution for a point-like source with an X-ray spectrum given by the core of \PSO (see the next subsection). In order to simulate the point spread function (PSF) of the core alone, we used the Model of AXAF Response to X-rays (\texttt{MARX}; version 5.5) Chandra end-to-end science simulator through the \texttt{CIAO} task \texttt{simulate\_psf}. In particular, we averaged the simulated PSF over 1000 trials with the core’s spectrum normalised at 100 times its actual flux as input in order to increase the statistics. From the comparison of the expected PSF and the observed radial distribution, it is clear that up to $\sim$2.5\arcsec \, \PSO is consistent with being a point-like source, whereas at larger radii there is a statistically significant abundance of observed X-ray photons (see Fig. \ref{fig:radial_prof}, top). In particular, the observed number of photons enclosed in each annulus is five to ten times larger than what is expected from a point source alone.

At the same time, by computing the angular profile of the photons detected in 20 sectors with an inner and outer radius of 2--10\arcsec, we find that the observed extended X-ray emission has an angular aperture of $\sim$50~deg centred at a P.A. of 55~deg. Interestingly, the number of photons observed in the opposite direction, where the counter-jet is expected, is consistent with the background level, thus confirming that we are observing the source at an angle close to the jet's axis. Based on the distributions reported in Fig. \ref{fig:radial_prof}, throughout the paper we consider the jet to have a radial and angular extension of 2.5\arcsec (from $r$=2.5\arcsec to $r$=5\arcsec) and 4\arcsec \, (from P.A.=25~deg to P.A.=80~deg), respectively. Assuming a viewing angle of $\theta_\mathrm{v} \sim 20^{\mathrm{o}}$ (see the next section), this projected distance, 15-30~kpc at \textit{z}$=$6.1, corresponds to a de-projected size of 40-90~kpc.

\subsubsection{X-ray spectral analysis}
Based on the values reported in the previous section, the region we adopted for the X-ray spectral analysis is reported with a solid blue ellipse in Fig. \ref{fig:image_XR} and contains a total of 27 photons between 0.5 and 7~keV.

\begin{figure}
   \centering
   \includegraphics[width=\hsize]{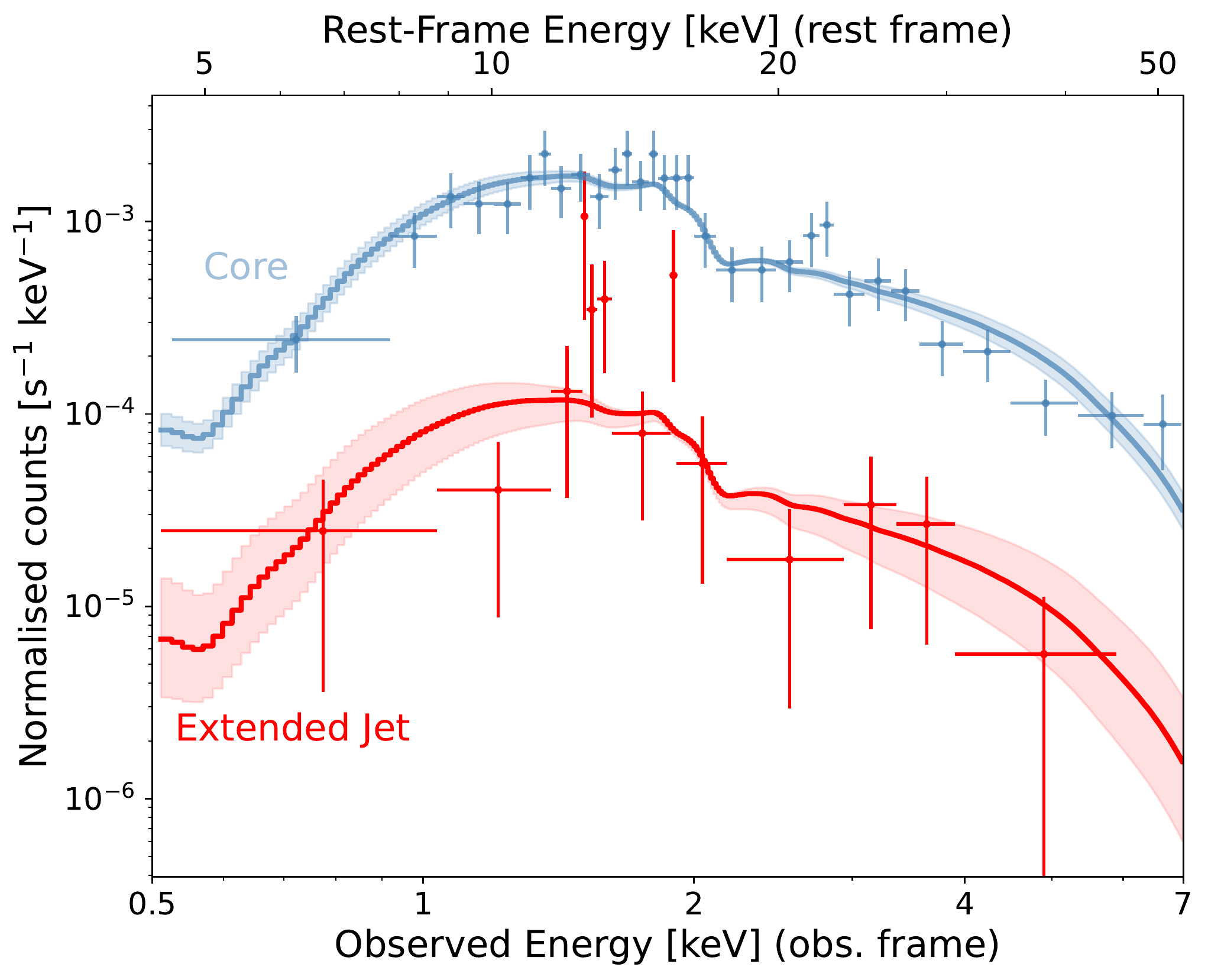}
      \caption{Folded X-ray spectrum of the core (top, in blue) and of the extended jet (bottom, in red) of PSO~J0309+27. Data are binned at a 2$\sigma$ significance for display purposes only. In both cases data were fitted with a Galactic absorbed power law. The best-fit spectra are reported as solid lines, and the shaded areas represent their 90\% confidence uncertainties.}
         \label{fig:spectra}
   \end{figure}

The analysis of the extended jet was performed in a similar way to the one of the core described in \cite{Moretti2021}. Using the \texttt{SPECEXTRACT} task, we extracted the core and jet counts from the two regions reported in Fig. \ref{fig:image_XR}, while for the background we considered an annulus between 12\arcsec and 25\arcsec centred at the position of the AGN.
We then analysed the extracted spectra using the \texttt{XSPEC} (v12.11.1) package and performed a fit to both the core and the jet emission in the energy range 0.5-7~keV (in order to reduce the background noise) with a power law absorbed by the Galactic column density along the line of sight (N$_H$=1.13$\times$10$^{21}$ cm$^{-2}$; \citealt{HI4PI2016}) and adopting the C-statistic \citep{Cash1979}. We report in Table \ref{Tab:AnalisiX} the results of the analysis. 
In the case of the core analysis, the `flaring' intervals described in \cite{Moretti2021} have not been considered. In Fig. \ref{fig:spectra} we report the X-ray spectra of the core and the extended jet of \PSO modelled as described above.

\
As already noted by \cite{Belladitta2020}, the X-ray properties of the core ($\Gamma_{\rm core}$=1.65 and $\alpha_{ox}$=1.13, $\tilde{\alpha}_{ox}$=1.02\footnote{Defined as: ${\alpha}_{ox}$=$-$0.384 log$\frac{L_\mathrm{2keV}}{L_\mathrm{2500\AA}}$ and  $\tilde{\alpha}_{ox}$=$-$0.303  log$\frac{L_\mathrm{10keV}}{L_\mathrm{2500\AA}}$.}) suggest that the observed X-ray emission is dominated by the beamed radiation produced within the jet oriented close to the line of sight (i.e. it is likely a blazar; e.g. \citealt{Ghisellini2015c, Ighina2019}). 
The overall X-ray-to-radio ratio ($L_\mathrm{2-10keV}$/$L_\mathrm{1.4GHz}$=2.76) is also consistent with the trend of increasing X-ray luminosities as a function of redshift observed up to \textit{z}$\sim$5.5 in \cite{Ighina2021}, which may be attributed to an increase in the total X-ray emission (resolved and not) due to the IC/CMB interaction.

The off-nuclear X-ray emission presents a relatively flat slope ($\Gamma_\mathrm{jet}$=1.79), which suggests that the population of electrons has not yet suffered significant energy losses \citep[e.g.][]{Achterberg2001}. Its rest-frame luminosity, L$_\mathrm{2-10keV}$=5.91$\times$10$^{44}$~erg~sec$^{-1}$, corresponds to $\sim$8\% of the core luminosity, making it one of the least core-dominated AGNs detected in X-rays (e.g. \citealt{Snios2021}) and at the same time one of the most X-ray-luminous extended jets observed to date (e.g. \citealt{McKeough16}).

\begin{threeparttable}
\centering
\caption{Best-fit parameters derived from the X-ray analysis of both the jet and core component shown in Fig. \ref{fig:image_XR}. In both cases a simple power law with only Galactic absorption is assumed (N$_H$=1.13$\times$10$^{21}$ cm$^{-2}$; \citealt{HI4PI2016}).}
\label{Tab:AnalisiX}
\setlength{\tabcolsep}{4pt}
\begin{tabular}{lcccc}
    \hline
    \hline
    &   $\Gamma$    &   f$_\mathrm{0.5-7keV}$ &   L$_\mathrm{2-10keV}$     & cstat / d.o.f.\\
    & & {\small 10$^{-15}$~erg~s$^{-1}$~cm$^{-2}$} & {\small 10$^{44}$~erg~s$^{-1}$} &\\
    \hline
    \\
    Core:   &    1.65$^{+0.18}_{-0.18}$ & 47.82$^{+1.21}_{-0.82}$ & 78.33 $^{+17.02}_{-14.01}$&   128.9 / 176 \\
    \\
    Jet:    &  1.79$^{+0.74}_{-0.69}$ & 3.08$^{+0.49}_{-0.30}$ & 5.91 $^{+6.81}_{-3.20}$&   26.9 / 20  \\
\\
\hline
\hline
\end{tabular}

\end{threeparttable}

\section{Discussion}
\label{sec:discussion}

 The mechanism responsible for the high-energy emission observed in large-scale extragalactic jets is still debated. The same electrons emitting in the radio band via the synchrotron process can be responsible for the observed X-ray flux via IC with the CMB photons \citep[e.g][]{Schwartz2019}. However, this would require that most jets remain relativistic on scales of up to 10-100~kpc \citep[e.g.][]{Tavecchio2000}. As a consequence, one would also expect this same interaction to produce bright $\gamma$-ray emission. The non-detection of such radiation with the \textit{Fermi} Large Area Telescope (\textit{Fermi}/LAT) therefore rules out the IC/CMB interaction as the main mechanism responsible for the X-ray radiation produced in the majority of extended jets in the local Universe (e.g. \citealt{Meyer2014, Meyer2015}). Therefore, another process must be taking place to account for the observed X-ray emission of these jets. The most plausible solution is the presence of a second population of electrons, in addition to the one responsible for the synchrotron radio and IC/CMB X-ray emissions. This population would have to be accelerated to very high energies ($\gamma \sim 10^8-10^9$ or $\sim$100~TeV for magnetic fields of the order of a few microgauss) and then emit in X-rays through the synchrotron process \citep[e.g.][]{Atoyan2004,Kataoka2005}. How these electrons can be accelerated up to very high energies far from the jet launch site is not fully understood yet. In the following we consider the scenario detailed in \cite{Tavecchio2021}.

\subsection{Two electron population model}

The model adopted here for the emission of very high-energy electrons (presented in \citealt{Tavecchio2021}) is based on the `shear acceleration' mechanism (e.g. \citealt{Rieger2019}), in which electrons can be accelerated to ultra-relativistic energies by magnetic turbulence in a shear layer that surrounds the main body of the jet (the spine). This shear layer is expected to be characterised by a strong radial gradient of the velocity. In these conditions, particles diffusing in the layer experience a continuous energy gain through scattering by the turbulence moving at different speeds. An important feature of this model is that the acceleration can only proceed at a sufficiently fast rate in the case of highly energetic electrons. Therefore, this mechanism requires a process able to pre-accelerate the electrons. Possible candidates for the pre-acceleration include turbulence in the flow \citep{Liu2017} or magnetic reconnection triggered by instabilities at the jet boundary \citep{Sironi2021}. \cite{Tavecchio2021} assumes that electrons experiencing shear acceleration are those accelerated by shocks along the jet (i.e. the ones responsible for the synchrotron emission at low frequency).

Even with a pre-acceleration mechanism, the shear acceleration timescale is relatively long. This feature gives us a simple explanation for the phenomenology observed at both low and high redshift. In fact, at low redshift the relatively small radiative losses allow the electrons in the shear layer to be accelerated up to high energies and therefore to emit in  X-rays through a synchrotron process.
At high redshift, where the CMB energy density is much higher, the situation changes since the IC/CMB cooling time  is significantly shorter than the shear acceleration timescale, even for the most energetic electrons. This severely hampers the acceleration and effectively limits the contribution of the second population to the X-ray emission.
At the same time, the increased CMB energy density determines a luminous IC/CMB emission from the low-energy electrons accelerated at the shock, which naturally accounts for the bright X-ray emission. In this scheme it is therefore natural to expect the second electron population to contribute to or even dominate the X-ray emission of low-$z$ jets, while at high redshift the main contribution derives from the IC/CMB emission of the low-energy electron component.

\begin{table*}
\centering
\begin{tabular}{ccccccccccc}
\hline
\hline
 $\Gamma_b$ & $\theta$ & $\gamma_{\rm cut}$ $(\times 10^5)$ & $K$ & $n_{\rm sh}$ & $B$ & $\delta$ & $R$ &$\tau_{\rm inj}$ & $t$ & $P_{\rm jet}$\\
 \, [1] & [2] & [3] & [4] & [5] & [6] & [7] & [8] & [9] & [10] &[11]  \\
\hline
 1.7 & 20 & 4 & 0.45 & 2.5 & 13 &2.5 & $2\times 10^{21}$& $5\times 10^2$ & 10 & 2.2\\
\hline
\hline
\end{tabular}
\vskip 0.4 true cm
\caption{
Parameters of the model.
[1]: Jet bulk Lorentz factor;
[2]: viewing angle (deg);
[3]: cutoff electron Lorentz factor of the shock component;
[4]: normalisation of the shock electron energy distribution (particle cm$^{-3}$);
[5]: slope of the shock electron energy distribution;
[6]: magnetic field ($\mu$G);
[7]: Doppler factor;
[8]: jet radius (cm);
[9]:  injection timescales for the shear acceleration in units of the light-crossing time, $r_j/c$ (where $r_j$ is the radius of the jet);
[10]: time in units of the light-crossing time, $r_j/c$.
[11]: jet power ($10^{46}$ erg s$^{-1}$).}
\label{tab:param}
\end{table*}

\subsection{Application to PSO~J0309+27}

PSO~J0309+27, with its powerful jet at an unprecedented high redshift, represents the ideal laboratory for testing the scenario presented above. In order to check the consistency with theoretical expectations, we compared our data to the two populations of electrons model used by \cite{Tavecchio2021} to reproduce the emission of the prototypical extended jet associated with the blazar PKS~J0637$-$752 at $z$=0.65.
In principle, this model is characterised by many free parameters. However, constraints from jet energetics and acceleration efficiency restrict the range of several parameters. In particular, the magnetic field intensity is constrained by the total energy and cooling of the electrons around 10~$\mu$G.
Since the treatment of the acceleration is performed in the non-relativistic limit, we assume a relatively small bulk Lorentz factor, $\Gamma_{\rm b}=1.7$. 

\begin{figure}
   \centering
    \includegraphics[width=\hsize]{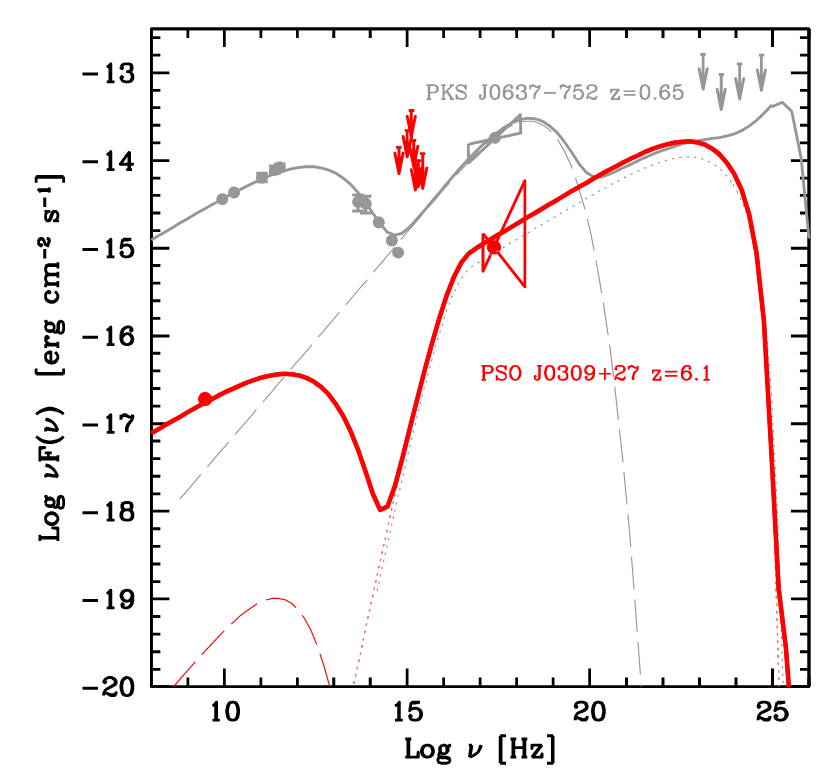}
      \caption{Observed SED of the extended emission of the source PKS~J0637$-$752 at $z$=0.65 (grey points), based on which the physical parameters of the model have been set (grey lines). The red lines show the same model but at the redshift of PSO~J0309+27, $z$=6.1. The X-ray, optical-to-NIR, and radio measurements or upper limits of the extended jet emission in \PSO are shown as red points and arrows, and those of PKS~J0637$-$752 are shown as grey points and arrows. In grey we also show the upper limits from \textit{Fermi}/LAT for both sources. We stress that the only difference between the input parameters of the two curves is the redshift. At low $z$ the extended X-ray emission is dominated by the synchrotron component (dashed lines) of the second population of electrons accelerated up to ultra-high energies ($\gamma \sim 10^{8-9}$). At high $z$, on the other hand, electrons are cooled down by the IC/CMB interaction much faster than they can be accelerated, and thus the IC/CMB emission (dotted lines) dominates in the X-rays and the synchrotron emission is significantly quenched.}
         \label{fig:multi_SED}
   \end{figure}

Remarkably, by adopting the same parameter values for the jet characteristics (listed in Table \ref{tab:param}) that reproduce the emission of PKS~J0637$-$752 at $z$=0.65 (solid grey line in Fig. \ref{fig:multi_SED}) and only varying the level of the energy density of the CMB (a factor of $\sim$350 increase from $z$=0.65 to $z$=6.1), the same model can also nicely fit the \PSO data (solid red line in Fig. \ref{fig:multi_SED}).
In the case of PKS~J0637$-$752, the radio and X-ray humps are produced by the synchrotron emissions of the low- and high-energy electron population, respectively. The contribution from the IC/CMB emission of the low-energy electrons is almost negligible (dotted grey lines) up to the hard X-ray band. In the $\gamma$-ray band, the model predicts a component due to the combination of synchrotron self-Compton and IC/CMB emission of the high-energy component. However, the absorption of the gamma rays by the extragalactic background light at optical-UV frequencies determines a narrow cutoff around 50 GeV ($\sim$10$^{25}$~Hz).
At high redshift (i.e. PSO~J0309+27; red lines), the increasing level of the CMB energy density determines the rapid cooling of electrons with Lorentz factors exceeding $\gamma\sim 10^5$. In these conditions the shear acceleration process is basically inactive, while the IC/CMB emission from the low-energy electron component nicely matches the observed X-ray emission. 

We note that the relatively small bulk Lorentz factor adopted here is not in contrast with the larger values derived by \cite{Spingola2020} and \cite{Moretti2021} when studying the innermost regions of the jet of PSO~J0309+27 in the radio and X-ray band, respectively. This is not unexpected since, at very large distances from the central engine (\textgreater\ 1 kpc), the jet likely decelerates \citep[e.g.][]{Bridle1999,Asada2012}. Otherwise, large bulk Lorentz factors ($\Gamma_b \gtrsim 5-10$) would imply a strong IC/CMB $\gamma$-ray emission detectable by the \textit{Fermi}/LAT, which has not been observed in the majority of the extended jets in the local Universe \citep[e.g.][]{Meyer2015}. 

\begin{table*}
\centering
\caption{X-ray and radio properties of the extended jet  of PSO~J0309+27 extracted from the \textit{Chandra} and VLASS images. The errors are reported at 90\% confidence.}
\label{Tab:Jet}

\begin{tabular}{lcccccc}
\hline
\hline
   S$_\mathrm{1keV}$ (nJy) &    S$_\mathrm{3GHz}$ (mJy) &   $\alpha_\mathrm{rx}$ & $\nu_\mathrm{X} S_\mathrm{X}$/$\nu_\mathrm{r} S_\mathrm{r}$ & (L$_\mathrm{jet}$/L$_\mathrm{core}$)$_\mathrm{X}$ & (L$_\mathrm{jet}$/L$_\mathrm{core}$)$_\mathrm{r}$  \\
\hline

 0.42$^{+0.21}_{-0.18}$ &   0.63$\pm$0.20 &  -0.78$\pm$0.03  & 54$^{+32}_{-29}$ &  8\% & 5\% \\
\hline
\hline
\end{tabular}
\end{table*}

\subsection{Redshift evolution of the extended X-ray emission}

The two cases analysed in the previous subsection represent two extremes of a continuous sequence regulated by the CMB energy density (i.e. redshift). In order to show the dependence of the kiloparsec-scale jet emission as a function of redshift and to facilitate the comparison with other studies on extended jets, we adopted the X-ray-to-radio flux ratio ($\nu_{\rm X} S_{\rm X}$/$\nu_{\rm r} S_{\rm r}$) as well as 
the corresponding $\alpha_{\rm rx}$ parameter, defined as the slope of a power law connecting the radio and X-ray emission in the observed frame (e.g. \citealt{Marshall2005}):
\begin{equation}
    \alpha_\mathrm{rx}= - \frac{\mathrm{log \, (S_X \, / \, S_r)}} {\mathrm{log \, (\nu_X \, / \, \nu_r)}} = 1 - \frac{{\rm log \, }(\nu_{\rm X} S_{\rm X}/\nu_{\rm r} S_{\rm r})}{{\rm log \, }(\nu_{\rm X}/\nu_{\rm r})}
.\end{equation}
In both cases, the values are taken at 1~keV and at 3~GHz in the observed frame. The X-ray and radio flux densities of \PSO were extracted from the \textit{Chandra} and VLASS images and are reported in Table \ref{Tab:Jet} together with the X-ray-to-radio flux and the jet-to-core (both X-ray and radio) ratios.

\begin{figure}
   \centering
   \includegraphics[width=\hsize]{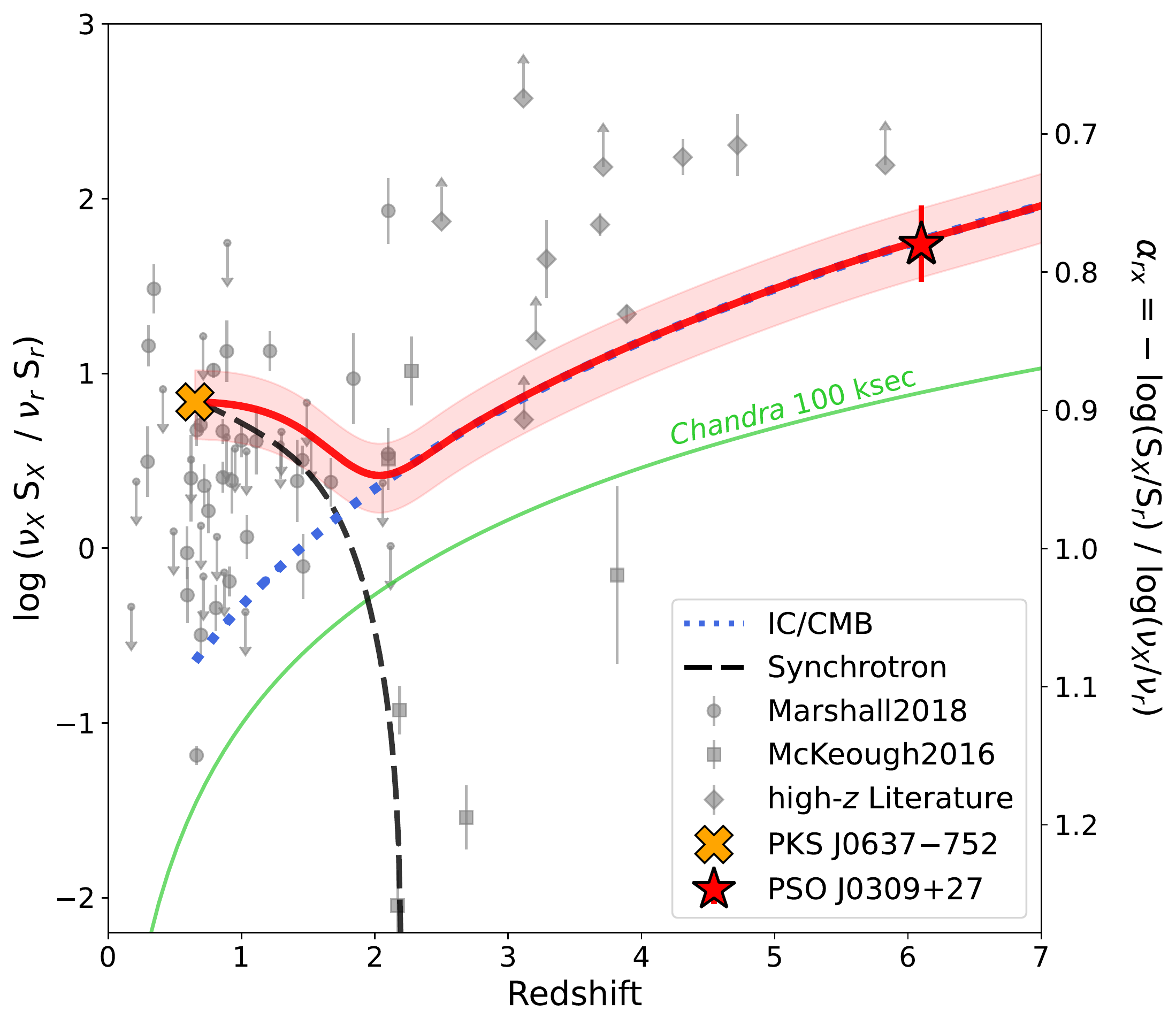}
      \caption{X-ray-to-radio ratios (and $\alpha_\mathrm{rx}$) of the extended jets observed by \textit{Chandra} reported in the literature. The data are from: \cite{Marshall2018}, circles at \textit{z}$\lesssim$2; \cite{McKeough16}, squares at 2$\lesssim$\textit{z}$\lesssim$4.7 (given by the sum of their `detected' components); \cite{Siemiginowska2003,Cheung2006,Simionescu2016,Schwartz2020,Worrall2020,Snios2021}; and \cite{Connor2021}, diamonds at \textit{z}$\gtrsim$3. If the same object is reported in more than one work, we consider the most recent.
      \PSO ($z$=6.1) is represented by a red star and PKS~J0637$-$752 ($z$=0.65) by an orange cross. The solid red line connecting the two sources is the evolution of the X-ray-to-radio ratio as a function of redshift based on the model reported in Fig. \ref{fig:multi_SED}, where the extended X-ray emission is given by the sum of a synchrotron (dashed black line) and an IC/CMB component (dotted blue line). The solid green line represents the limit above which an extended X-ray jet with the same radio luminosity as the one found in \PSO can be observed with a 100~ksec \textit{Chandra} exposure.}
      
         \label{fig:XR_jets}
   \end{figure}

In Fig. \ref{fig:XR_jets} we show the evolution of the X-ray-to-radio ratio based on the model described above as a function of redshift (solid red line). As explained in the previous subsection, while at low redshift (PKS~J0637$-$752, orange cross) the high-energy synchrotron emission dominates the observed X-ray range, at high redshift (PSO~J0309+27, red star) the only mechanism able to efficiently produce the observed extended X-ray emission is the IC/CMB interaction. Based on the parameters adopted here, the transition between the two mechanisms is expected to be around \textit{z}$\sim$2. Beyond this value the peak of the high-energy synchrotron emission shifts below the X-ray band due to cooling, and therefore its contribution to the total radiation observed is negligible (dashed black line). At the same time, the increase in the CMB energy density boosts the IC/CMB emission, which becomes dominant at \textit{z}\textgreater2 (dotted blue line), where we expect to observe its typical evolutionary trend $\propto$(1+$z$)$^{3+\alpha}$ (e.g. \citealt{Worrall2009}). This is consistent with the other detailed studies on extended jets at \textit{z}$\sim$3-4 \citep[e.g.][]{Cheung2012,Worrall2020}. 

In Fig. \ref{fig:XR_jets} we also compare the relative X-ray intensity of the extended emission with respect to the radio emission for \PSO to the other resolved jets in AGNs at lower redshift available in the literature. All the comparison sources are quasars, as is PSO~J0309+27. If the radio data were taken at a different frequency with respect to PSO~J0309+27, we computed the corresponding flux density at 3~GHz assuming a radio spectral index of $\alpha_\mathrm{r}$=1, as is typically observed in spatially resolved jets at kiloparsec scales \citep[e.g.][]{Cheung2006,Cheung2012}. The data are  largely scattered, probably reflecting the wide variety of the physical properties of the jets and their components. As noted by previous authors (e.g. \citealt{McKeough16,Marshall2018}), although low values of the X-ray-to-radio ratio are not observed at high redshift, the correlation with redshift is not highly significant. In particular, the sources at $z$\textgreater2 do not belong to a single complete sample, and the values reported in the literature are often biased towards objects with a strong X-ray jet since many of them are detections resulted from very short \textit{Chandra} observations ($\sim$10~ksec; e.g. \citealt{Snios2021}). As a reference, we also show  the limit above which extended jets with similar radio properties to \PSO (i.e. radio luminosity)  can be detected with a relatively deep (100~ksec) \textit{Chandra} observation (solid green line).

\section{Summary and conclusions}
\label{sec:conclusions}

In this work we have presented the X-ray detection and analysis of the most distant kiloparsec-scale? jet that is spatially resolved, at $z$=6.1. From dedicated \textit{Chandra} observations we found that the X-ray emission extends up to about 4\arcsec, which is equal to a de-projected linear size of $\sim$70~kpc (assuming $\theta_v\sim20^{\rm o}$). This kiloparsec-scale jet is characterised by a high luminosity (L$_{\rm 2-10keV}$=5.9$^{+6.8}_{-3.2} \, \times$10$^{44}$~erg~s$^{-1}$) as well as one of the largest core-to-jet ratios (8\%). Moreover, from the detection of this extended component in the VLASS radio survey, we were also able to constrain its SED and its X-ray-to-radio ratio. The very high redshift of \PSO makes it an ideal source for testing the redshift evolution of the emission in extended jets expected from the IC/CMB trend. 

By comparing the X-ray-to-radio flux ratio of \PSO to the other extended jets analysed in the literature, we found that the value derived for \PSO is larger than what is typically measured in the local Universe (e.g. \citealt{Marshall2018}), as expected from the IC/CMB model.  Indeed, assuming physical parameters with values typically seen in the local Universe and only mild relativistic boosting ($\Gamma_{\rm b}\sim$1.7), the multi-wavelength emission of the \PSO jet is fully consistent with the IC/CMB interaction of the same electrons responsible for the synchrotron radiation in the radio band. 

Although similar conclusions were drawn for a few sources at \textit{z}$\sim$3-4 (e.g. \citealt{Worrall2020}), we also argue that the IC/CMB interaction is the only mechanism able to efficiently produce the extended X-ray emission observed at these redshifts. Indeed, we found that the expected contribution to the extended X-ray emission from a second population of electrons (modelled as discussed in \citealt{Tavecchio2021}) is naturally quenched at high redshift since the time needed by the electrons to be accelerated up to very high energies becomes longer than the IC/CMB cooling time. In particular, we have shown how the high-energy synchrotron model can reproduce the radio-to-X-ray emission in jets up to \textit{z}$\sim$2, after which the redshift dependence of the CMB energy density has a double effect on the two populations of electrons: on the one hand, it enhances the X-ray (and $\gamma$-ray) radiation produced through the IC/CMB interaction by the low-energy population; on the other hand, it more efficiently cools the most energetic electrons, which, in turn, cannot be accelerated up to the energies needed to produced synchrotron X-ray emission.
Despite the general consistency of this picture with the observations collected so far, detailed studies of statistical samples of high-$z$ jets (with both radio and X-ray data) need to be performed in order to put this result on a more solid basis.

\begin{acknowledgements}
We want to thank G. Ghisellini for his helpful comments to the article. We also thank the anonymous referee for their comments.
We acknowledge financial contribution from the agreement ASI-INAF n. I/037/12/0 and n.2017-14-H.0 and from INAF under PRIN SKA/CTA FORECaST. CS acknowledges financial support from the Italian Ministry of University and Research - Project Proposal CIR01\_00010.\\
The scientific results reported in this article are based to a significant degree on observations made by the Chandra X-ray Observatory.\\
This research has made use of software provided by the Chandra X-ray Center (CXC) in the application packages CIAO, ChIPS, and Sherpa.\\
This research made use of Astropy (\url{http://www.astropy.org}) a community-developed core Python package for Astronomy \citep{astropy2018}.

\end{acknowledgements}

\bibliographystyle{aa} 
\bibliography{referenze} 

\end{document}